\documentclass[12pt]{article}

\usepackage{amssymb}
\usepackage{amsmath}
\usepackage{latexsym}

\catcode`@=11
\renewcommand{\section}{\@startsection{section}{1}{0pt}{\medskipamount}
{\medskipamount}{\large\bf}}
\numberwithin{equation}{section}
\catcode`@=12

\def\beq{\begin{eqnarray}}    
\def\eeq{\end{eqnarray}}      

\def\ln{\,\mbox{ln}\,}                  
\def\tr{\,\mbox{tr}\,}                  
\def\det{\,\mbox{det}\,}                
\def\sDet{\,\mbox{sDet}\,}              

\def\pa{\partial}                       

\def\={\ =\ }

\begin{document}

\begin{center}

{\large\bf A systematic study of finite BRST-BFV transformations \\
in $Sp(2)$-extended generalized Hamiltonian formalism}

\vspace{18mm}

{\Large Igor A. Batalin$^{(a)}\footnote{E-mail:
batalin@lpi.ru}$\;,
Peter M. Lavrov$^{(b, c)}\footnote{E-mail:
lavrov@tspu.edu.ru}$\;,
Igor V. Tyutin$^{(a)}\footnote{E-mail: tyutin@lpi.ru}$
}

\vspace{8mm}

\noindent ${{}^{(a)}}$
{\em P.N. Lebedev Physical Institute,\\
Leninsky Prospect \ 53, 119 991 Moscow, Russia}

\noindent  ${{}^{(b)}}
${\em
Tomsk State Pedagogical University,\\
Kievskaya St.\ 60, 634061 Tomsk, Russia}

\noindent  ${{}^{(c)}}
${\em
National Research Tomsk State  University,\\
Lenin Av.\ 36, 634050 Tomsk, Russia}

\vspace{20mm}

\begin{abstract}
\noindent
We study systematically finite BRST-BFV transformations in $Sp(2)$-extended generalized
Hamiltonian formalism. We present explicitly their Jacobians and the form of a solution
to the compensation equation determining the functional field dependence of finite
Fermionic  parameters, necessary to generate arbitrary  finite change of gauge-fixing
functions in the path integral.
\end{abstract}

\end{center}

\vfill



\section{INTRODUCTION}

In our recent articles \cite{BLT-BRST-BFV,BLT-BRST-BV}, we have
studied finite BRST transformations in the  framework of the
generalized Hamiltonian (BFV) formalism \cite{FV,BVhf,FF}, as well
as of the field-antifield (BV) formalism \cite{BV,BV2}. An important
result was obtained  that a finite BRST transformation was capable
of generating an arbitrary finite change of gauge-fixing conditions
in these quantization methods. Both these quantization schemes are
essentially based on the BRST symmetry principle \cite{brs1,brs2,t}.
In addition to usual BRST symmetry, the anti-BRST symmetry was also
known \cite{CF,Oj} for Yang-Mills theories in special gauges. For a
long time,  an opinion was dominating  that  Hamiltonian
quantization of dynamical systems with constraints, preserving the
BRST-anti-BRST (or extended BRST) symmetry, could  be performed only
if  structure coefficients of a gauge algebra were independent of
phase variables \cite{Hw}, and in turn, the Lagrangian quantization
respecting  the extended BRST symmetry was possible only for gauge
theories with closed algebra \cite{Sp}. Later,  quantization methods
were discovered based essentially on the  extended BRST symmetry
principle in the Hamiltonian formalism of arbitrary dynamical
systems with constraints \cite{BLTh1,BLTh2,BLTh3}, as well as in the
Lagrangian formalism of general gauge theories
\cite{BLT4,BLT5,BLT6}. As the variables of extended phase space in
the Hamiltonian formalism, as well as the variables of extended
configuration space in the Lagrangian formalism, form irreducible
representations of the $Sp(2)$ group, these methods are labeled with
abbreviation $"Sp(2)"$.

In this paper,  we will extend the results of our previous paper
\cite{BLT-BRST-BFV} to the case of $Sp(2)$ generalized Hamiltonian
formalism \cite{BLTh1,BLTh2,BLTh3}, with two Fermionic parameters,
so that the global $Sp(2$) symmetry is included by construction.
Physically, that $Sp(2)$ symmetry establishes the "democracy"
between ghosts and anti-ghosts. Notice that in the $Sp(2)$ symmetric
formalism, a gauge-fixing function is a Boson, in contrast to the
standard  case \cite{FV,BVhf,FF}.  At the same time, the $Sp(2)$
vector-valued BRST-BFV  generators enter the unitarizing Hamiltonian
quadratically. We define effectively the $Sp(2)$-extended BRST-BFV
transformations by the corresponding Lie equations in the "plane" of
the two Fermionic parameters. In principle, our new construction
follows the same general logic as we did in our previous article
\cite{BLT-BRST-BFV}. The main new feature is that the Jacobian of
the $Sp(2)$-extended BRST-BFV transformation is expressed in terms
of  a determinant of a $2\times2$ matrix, so that the compensation
equation becomes $2\times2$ matrix-valued as well.
\\

\section{Sp(2)-EXTENDED FINITE BRST-BFV TRANSFORMATIONS AND THEIR
JACOBIANS}

Let
\beq \label{E2.1}
z^i=(q;p),\qquad \varepsilon(z^i)=\varepsilon_i       
\eeq
be a complete set of canonical variables specific to the  phase space of
$Sp(2)$-extended generalized Hamiltonian formalism. We proceed with the
following path integral representation for the partition function $Z_{F}$,
\beq \label{E2.2}
Z_{F} = \int Dz \exp\left[\left(\frac{i}{\hbar}\right)W_F\right],                                       
\eeq where the action $W_F$ is defined as (all the $Sp(2)$ indices
such as $"a,b,..."$ take the two values $a=1,2$; by $F$ we denote a
gauge-fixing Boson in the $Sp(2)$-extended formalism), \beq
\label{E2.3}
W_F=\int\left[\left(\frac{1}{2}\right)z^i(t)\omega_{ik}\dot{z}^k(t)-H_F(t)\right]dt\;,
\eeq 
\beq
\label{E2.4}
H_F(t)=\mathcal{H}(t)+\left(\frac{1}{2}\right)\varepsilon_{ab}\{\{F,\Omega^b\},\Omega^a\}_t\;,    
\eeq
\beq
\label{E2.5}
\{\Omega^{a},\mathcal{H}\}=0\;, \qquad    \{\Omega^{a},\Omega^{b}\}=0\;,              
\eeq
\beq
\varepsilon(\mathcal{H})=\varepsilon(F)=0\;,\qquad\varepsilon(\Omega^{a})=1\;,     
\eeq
\beq
\mathcal{H} = H_0+...\;,\qquad \Omega^{a} = c^{a}T+...\;.                
\eeq
Here in (\ref{E2.3}), $z^i(t)$ are functions of time (trajectories),
$\dot{z}^k(t)=dz^k(t)/dt$, $H_F(t)$, $\mathcal{H}(t)$, $\Omega(t)$, $F(t)$
are local functions of time: $H_F(t)=\left.H_F(z)\right|_{z\rightarrow z(t)}$
and so on, $\{,\}_t$ means the Poisson superbracket for fixed time $t$:
$\{F,\Omega^a\}_t=\left.\{F(z),\Omega^a(z)\}\right|_{z\rightarrow z(t)}$
and so on, $\{z^i,z^k\}=\omega^{ik}=\hbox{const}=-\omega^{ki}
(-1)^{\varepsilon_i\varepsilon_k}$ is an invertible even matrix;
$\omega_{ik}=\omega_{ki}(-1)^{(\varepsilon_i+1)(\varepsilon_k+1)}$
stands for an inverse to $\omega^{ik}$; $\varepsilon^{ab}$ is the constant
$Sp(2)$-invariant antisymmetric tensor, while $\varepsilon_{ab}$ stands for an
inverse to $\varepsilon^{ab}$, $\varepsilon^{ab}\varepsilon_{bc}=\delta^a_c$,
$\mathcal{H}$ and $F$ are Bosons while $\Omega^a$ are Fermions. It follows from
(\ref{E2.4}) that
\beq
\{\Omega^a,H_F\}=0.       
\eeq

We define finite BRST-BFV transformations in their differential form, by the
following Lie equations in the "plane" of their Fermionic parameters $\mu_{a}$,
\beq
\label{E2.9}
&&\overline{z}^i(z,\mu)\overleftarrow{\pa}^a=
\{\overline{z}^i,\overline{\Omega}^a\}_{\overline{z}},\quad
\overleftarrow{\pa}^a=\frac{\overleftarrow{\pa}}{\pa\mu_a},    \\     
\label{E2.10}
&&\overline{\Omega}^a=\Omega^a(\overline{z}),\quad\overline{z}^i|_{\mu=0}=z^i.   
\eeq

It follows from (\ref{E2.9}), (\ref{E2.10}) that
\beq
\overline{z}^i(z,\mu)\overleftarrow{\pa}^a\overleftarrow{\pa}^b=
\{\{\overline{z}^i,\overline{\Omega}^a\}_{\overline{z}},
\overline{\Omega}^b\}_{\overline{z}}\;,  
\eeq
and then
\beq \label{E2.12}
\overline{z}^i=z^i+\{z^i,\Omega^a\}\mu_a+\frac{1}{2}\{\{z^i,\Omega^a\},
\Omega^b\}\mu_b\mu_a=z^i\exp[\{(...),\Omega^a\}\mu_a]\;.   
\eeq

The last equality in (\ref{E2.12}) does confirm explicitly the canonicity of
that transformation with $\mu_a=\hbox{const}$.

By applying the operators $\overleftarrow{\pa}^a$ to
$\overline{\Omega}^a$ and $\overline{\mathcal{H}}$, and using the Lie
equations (\ref{E2.9}) together with (\ref{E2.5}), we get
\beq
\overline{\Omega}^a=\Omega^a\;, \; \overline{\mathcal{H}} = \mathcal{H}\;.  
\eeq

In the same way, we get for the unitarizing Hamiltonian $H_F$ (\ref{E2.4})
\beq \label{E2.14}
\overline{H_F} = H_{F}\;.                          
\eeq

The finite BRST-BFV transformations of trajectories $z^i(t)$ have the form
\beq \label{E2.12a}
\overline{z}^i(i)=z^i(t)+\{z^i,\Omega^a\}_t\;\mu_a+\frac{1}{2}\{\{z^i,\Omega^a\},
\Omega^b\}_t\;\mu_b\;\mu_a\;.   
\eeq
In general, the two Fermionic parameters in rep. (\ref{E2.12a}) are allowed to
be integral functionals $\mu_a=\mu_a[z]$ of the whole trajectory $z(t)$,
$-\infty<t<\infty$. However, by themselves, $\mu_a$ are independent of the
current time $t$ and local position $z$,
\beq
d_t\mu_a[z]=0, \quad \pa_i\mu_a[z] =0,                     
\eeq
where we have denoted
\beq
d_t=\frac{d}{dt}, \quad   \pa_i=\frac{\pa}{\pa z^i}.              
\eeq
Thus, only a functional derivative such as $\delta/\delta z(t)$ is capable to
differentiate the $\mu_a[z]$ nontrivially.

By applying the operators $\overleftarrow{\pa}^a$ to the kinetic
part of the action (\ref{E2.3}) one and two times, and using the Lie
equation (\ref{E2.9}), we get for the total action (\ref{E2.3})
\beq
\label{E2.15} \overline{W_F}=W_F+\frac{1}{2}\left[(N -2)\Omega^a\mu_a+
\frac{1}{2}\{(N\Omega^a),\Omega^b\}\;\mu_b\;\mu_a\right]_t\Big|_{-\infty}^{+\infty},    
\eeq
where $N =z^i\pa_i$, and we have used (\ref{E2.14}). For the class of
trajectories whose asymptotic is such that the boundary term in the square
bracket in the r. h. s. in (\ref{E2.15}) is zero, the total action (\ref{E2.3}) is
invariant,
\beq \label{E2.16}
\overline{W_F}=W_F.                      
\eeq

Now, let us consider the functional Jacobian,
\beq
\nonumber
&&J=\sDet\left[\left(\overline{z}^i(t)
\frac{\overleftarrow{\delta}}{\delta z^j(t')}\right)\right]\!= \\
\label{E2.19}
&&=\sDet\left[\left(\overline{z}^i(z,\mu)
\overleftarrow{\pa}_j\right)_t\delta(t-t')+
\left(\overline{z}^i(z,\mu)\overleftarrow{\pa}^a\right)_t\!
\left(\mu_a[z]\frac{\overleftarrow{\delta}}{\delta z^j(t')}\right)\right]\!.      
\eeq

We factorize the Jacobian (\ref{E2.19}) in the form
\beq \label{E2.20}
J = J_1J_2\;,                     
\eeq
where
\beq
\label{E2.21}
J_1=\sDet[(G^{-1})^i_k(t,t';\lambda = 1)]\;,    
\eeq
\beq
\label{E2.22}
J_2=\sDet\left[\left(\overline{z}^i(z,\mu)
\overleftarrow{\pa}_j\right)_t\delta(t-t')\right]\;,
\eeq
\beq
\label{E2.23}
G^i_k(t,t'';\lambda)=\delta^i_k\delta(t-t'')-
\lambda\{z^i,\Omega^a\}_t A_{ak}(t'')\;, %
\eeq           
\beq
\nonumber
A_{ak}(t'')&=&\int dt'\mu_a[z]\frac{\overleftarrow{\delta}}{\delta z^j(t')}
G^j_k(t',t'';\lambda)=\\
\label{E2.24}
&=&[(1+\lambda\kappa)^{-1}]_a^b\left(\mu_b[z]
\frac{\overleftarrow{\delta}}{\delta z^k(t'')}\right)\;,  
\eeq
\beq
\label{E2.25}
\kappa_a^b=\mu_a[z]\int dt\frac{\overleftarrow{\delta}}{\delta z^i(t)}
\{z^i,\Omega^b\}_t\; .     
\eeq

Notice that the original explicit form of the integral equation for the Green
function $G$ is given by (\ref{E2.23}) with the first expression in
(\ref{E2.24}) standing for $A_{ak}(t'')$. Then, by multiplying by
$\mu_a[z][\overleftarrow{\delta}/\delta z^i(t)]$ from the left and taking the
$t$-integral, it follows a simple linear algebraic equation whose solution is
given by the second expression for $A_{ak}(t'')$ in (\ref{E2.24}).

It is a characteristic feature of the factor (\ref{E2.21}) that the operator
therein is nontrivial only for $\mu_a$ depending actually on fields. On the
other hand, in the factor (\ref{E2.22}), the corresponding operator has a
nontrivial part  proportional to undifferentiated $\mu_a$.
Let us consider the factors (\ref{E2.21}), (\ref{E2.22}) in more detail. For
the $J_{1}$ factor, we have
\beq
\nonumber
&&\ln J_1=\int_0^1d\lambda\int dtdt'G^i_j(t,t';\lambda)\{z^j,\Omega^a\}_{t'}
\left(\mu_a[z]\frac{\overleftarrow{\delta}}{\delta z^i(t)}\right)
(-1)^{\varepsilon_i}= \\
\nonumber
&&=-\int_0^1d\lambda\int dt'A_{aj}(t')
\{z^j,\Omega^a\}_{t'}= \\
\nonumber
&&=-\int_0^1d\lambda\int dt'[( 1 + \lambda \kappa )^{-1}]_a^b
\left(\mu_b[z]\frac{\overleftarrow{\delta}}{\delta z^i(t')}\right)
\{z^i,\Omega^a\}_{t'}= \\
&&=-\int_0^1\tr[(1+\lambda\kappa)^{-1}\kappa]d\lambda=-\tr[\ln(1+\kappa)]\;. 
\eeq

Thus, we have finally for $J_1$ (\ref{E2.21})
\beq \label{E2.27}
J_1=[\det(1+\kappa)]^{-1}.    
\eeq

Now, let us consider the ultra-local Jacobian $J_2$ (\ref{E2.22}).  As that
Jacobian involves only undifferentiated $\mu_{a}$, one is allowed to consider
the $\mu_a$ as a constant. On the other hand, in the latter case the
transformation is canonical, so that the Liouville theorem tells us that
\beq \label{E2.28}
J_2=1.    
\eeq

Indeed, by applying the operator $\overleftarrow{\pa}^a$ to $J_2$
and using the Lie equation (\ref{E2.9}), one can confirm the equality
(\ref{E2.28}) explicitly,
\beq
\nonumber
&&\left[\hbox{str}\ln\left(\overline{z}^i(z,\mu)\overleftarrow{\pa}_j
\right)\right]\overleftarrow{\pa}^a=
\left(z^i(\overline{z},\mu)\frac{\overleftarrow{\pa}}{\pa \overline{z}^k}\right)
\left(\overline{z}^k(z,\mu)\overleftarrow{\pa}_i\overleftarrow{\pa}^a\right)
(-1)^{\varepsilon_i}= \\
\label{E2.29}
&&=\{\overline{z}^k,\overline{\Omega}^a\}\frac{\overleftarrow{\pa}}{\pa \overline{z}^k}=
\omega^{kj}\frac{\pa}{\pa \overline{z}^j}
\frac{\pa}{\pa \overline{z}^k}\overline{\Omega}^a=0. 
\eeq

Now, we have:  $\ln J_2=\delta(0)\int dt\,\hbox{str}\ln[\overline{z}^i(z,\mu)
\overleftarrow{\pa}_j]_t$; as usual, we assume that the zero in
(\ref{E2.29}) is the principal factor. In that case, we arrive at (\ref{E2.28}).

So, from (\ref{E2.20}), (\ref{E2.27}), (\ref{E2.28}) we conclude finally
\beq \label{E2.30}
J = J_1=[\det(1+\kappa)]^{-1},   
\eeq
with $\kappa$ given by (\ref{E2.25}).
\\

\section{MATRIX-VALUED COMPENSATION EQUATION AND \\
ITS EXPLICIT SOLUTION   }

Now, we would like to use the Jacobian (\ref{E2.30}) to generate arbitrary
finite change $\delta F$ of the gauge Boson $F$ in the action (\ref{E2.3}),
\beq \label{E3.31}
F  \rightarrow F_{1}  = F +  \delta F                             
\eeq
Due to the invariance (\ref{E2.16}) of the action (\ref{E2.3}), we
have for the partition function (\ref{E2.2}) in the new variables
\beq \label{E3.2}
Z_F=\int D\overline{z}\exp\left[\left(\frac{i}{\hbar}\right)\overline{W_F}\right]=
\int DzJ\exp\left[\left(\frac{i}{\hbar}\right)W_F\right].                   
\eeq
Let us require the following condition to hold
\beq
\label{E3.3}
J =\exp\left[-\left(\frac{i}{\hbar}\right)\int dt\left(\frac{1}{2}\right)\varepsilon_{ac}
\{\{\delta F,\Omega^c\},\Omega^a\}_t\right].                  
\eeq
It follows then the gauge-independence property of the partition function,
\beq \label{E3.4}
Z_{F_1}=Z_F,                                     
\eeq
for arbitrary finite $\delta F$. We call the condition (\ref{E3.3})
"a compensation equation".
Due to the formula (\ref{E2.30}), it follows that (\ref{E3.3}) is
rewritten as
\beq \label{E3.5}
\tr\{[\ln(1+\kappa)]_a^b\}=\tr(x_a^b),                     
\eeq
where the matrix-valued functional $x_a^b$ is defined by
\beq \label{E3.6}
x_a^b=\left(\frac{i}{\hbar}\right)\int dt\frac{1}{2}
\varepsilon_{ac}\{\{\delta F,\Omega^{c}\},\Omega^{b}\}_t.        
\eeq
Now, let us require the matrix-valued counterpart to (\ref{E3.5}) to hold
\beq
\label{E3.7}
[\ln(1+\kappa)]_a^b = x_a^b.                              
\eeq
Due to (\ref{E2.25}), Eq. (\ref{E3.7}) is rewritten in a more detail as
\beq
\label{E3.8}
\mu_a[z]\int dt\frac{\overleftarrow{\delta}}{\delta z^i(t)}
\{z^i,\Omega^b\}_t=[\exp(x)-1]_a^b.                          
\eeq
That is a functional equation to determine $\mu_{a}$. There is an obvious
explicit solution to that equation
\beq
\label{E3.9}
\mu_a=\mu_a[z;\delta F]=\left(\frac{i}{\hbar}\right)[f(x)]_a^b
\int dt\left(\frac{1}{2}\right)\varepsilon_{bc}\{\delta F,\Omega^c\}_t,        
\eeq
where  $x_a^b$ is defined in (\ref{E3.6}), and
\beq \label{E3.10}
[f(x)]_a^b=[(\exp(x)-1)\,x^{-1}]_a^b.                                  
\eeq Functional operators in l. h. s. in (\ref{E3.8}), when applying
to the rightmost factor in (\ref{E3.9}), yield the $x$ of
(\ref{E3.6}) to cancel the factor $x^{-1}$ in (\ref{E3.10}). On the
other hand, these functional operators do annihilate $x$ itself due
to the Jacobi identity and the second in (\ref{E2.5}). Thus we have
confirmed explicitly the compensation equation (\ref{E3.8}) to hold.
In the first order in $\delta F$, explicit solution (\ref{E3.9})
takes the usual form
\beq
\label{E3.11} \mu_a[z;\delta
F]=\left(\frac{i}{\hbar}\right)\int dt\left(\frac{1}{2}\right)\varepsilon_{ac}
\{\delta F,\Omega^c\}_t+O((\delta F)^2).             
\eeq

\section{Sp(2)-EXTENDED FUNCTIONAL BRST-BFV \\
TRANSFORMATIONS FOR TRAJECTORIES}

It appears quite natural to make our considerations above more transparent
by introducing a concept of functional BRST-BFV transformations. Namely, let
us define a functional operator of the form
\beq
\label{E4.1}
\overleftarrow{d}^a=\int dt \frac{\overleftarrow{\delta}}{\delta z^{i}(t)}
\{z^i,\Omega^a\}_t. 
\eeq
It follows from the second in (\ref{E2.5}) that the Fermionic operators (\ref{E4.1})
super-commute among themselves,
\beq
\label{E4.2}
\varepsilon(\overleftarrow{d}^a)=1,  \quad
[\overleftarrow{d}^a,\overleftarrow{d}^b]=
\overleftarrow{d}^a\overleftarrow{d}^b+
\overleftarrow{d}^b\overleftarrow{d}^a=0, \quad
\overleftarrow{d}^a\overleftarrow{d}^b\overleftarrow{d}^c=0.         
\eeq
The transformation (\ref{E2.12}) of a trajectory $z^i(t)$ is rewritten
in terms of the operators (\ref{E4.1}) as
\beq
\label{E4.3}
\overline{z}^i(t) =z^i(t)\left(1 +\overleftarrow{d}^a\mu_a+
\left(\frac{1}{2}\right)\overleftarrow{d}^a\overleftarrow{d}^b\mu_b\mu_a\right).                 
\eeq
For trajectory-independent parameters $\mu_a$, rep. (\ref{E4.3}) is
transformed to the form
\beq \label{E4.3a}
\overline{z}^i(t) =z^i(t)\exp[\overleftarrow{d}^a\mu_{a}].   
\eeq
Thus, the operators (\ref{E4.1}) are functional BRST-BFV generators
at a trajectory.

Functional Jacobian (\ref{E2.30}) is rewritten in terms of the generator
(\ref{E4.1}) as
\beq \label{E4.4}
J=\big[\det[\delta_a^b+(\mu_a[z]\overleftarrow{d}^b)]\big ]^{-1}.                     
\eeq
Compensation equation (\ref{E3.8}) takes the form
\beq \label{E4.5}
\mu_a[z]\overleftarrow{d}^b=[\exp(x)-1]_a^b,                             
\eeq
where
\beq
\label{E4.6}
 x_a^b=\left(\frac{i}{\hbar}\right)\left(\frac{1}{2}\right)\varepsilon_{ac}
 \int dt(\delta F(t)\overleftarrow{d}^c\overleftarrow{d}^b).                
\eeq
Similarly to (\ref{E4.6}), the gauge-fixed unitarizing Hamiltonian
at a trajectory $z^i(t)$ is rewritten as
\beq
\label{E4.7}
H_F(t)=\mathcal{H}(t)+\left(\frac{1}{2}\right)\varepsilon_{ab}
F(t)\overleftarrow{d}^b\overleftarrow{d}^a.      
\eeq
Thus, we conclude that all the main objects in our considerations
can be expressed naturally in terms of the functional BRST-BFV generators
(\ref{E4.1}).

Finally, let us represent the equality in (\ref{E4.3}) in the form
\beq \label{E4.8}
\overline{z}^i(t)=z^i(t)\overleftarrow{T}(\mu),          
\eeq
where the T-operators are defined by
\beq
\label{E4.9}
\overleftarrow{T}(\mu)=1+\overleftarrow{d}^a\mu_a +
\left(\frac{1}{2}\right)\overleftarrow{d}^a\overleftarrow{d}^b\mu_b\mu_a.           
\eeq
Their commutators  have the form
\beq
\nonumber
&&[\overleftarrow{T}(\mu),\overleftarrow{T}(\mu')] =
\overleftarrow{d}^{a}[(\mu_a\overleftarrow{d}^b)\mu'_b+
(\mu_a\overleftarrow{d}^b\overleftarrow{d}^c)\mu'_c\mu'_b]-
(\mu\Leftrightarrow\mu')+ \\
\label{E4.10}
&&+\overleftarrow{d}^a\overleftarrow{d}^b
[(\mu_a\overleftarrow{d}^c)\mu'_b\mu'_c-\left(\frac{1}{2}\right)(\mu'_b\mu'_a
\overleftarrow{d}^c)\mu_c+\left(\frac{1}{2}\right)(\mu_b\mu_a\overleftarrow{d}^c
\overleftarrow{d}^e)\mu'_e\mu'_c]-(\mu\Leftrightarrow\mu').           
\eeq
That nonlinear open algebra looks essentially more complicated as
compared to the corresponding algebra in the standard case \cite{BLT-BRST-BFV}.
\\

\section{Sp(2) VECTOR-VALUED WARD IDENTITIES DEPENDENT OF BRST-BFV
PARAMETERS/FUNCTIONALS}

As we have defined finite BRST-BFV transformations, it appears quite natural
to apply them
immediately to deduce the corresponding modified version of the Ward identity.
We will do
that just in terms of functional BRST-BFV generators introduced in Sec. 4.

As usual for that matter, let us proceed with the external-source
dependent generating functional,
\beq
\label{E5.1} Z_{F}( \zeta, z^*, z^{**} )  =
\int Dz \exp\left[\left(\frac{i}{\hbar}\right) W_{F}( \zeta, z^*, z^{**} ) \right],                         
\eeq
\beq
\label{E5.2}
W_{F}( \zeta, z^*, z^{**} ) = W_{F}  + \int dt \left( \zeta_{i} z^{i} +
z^*_{ia} z^{i} \overleftarrow{d}^{a} +
z^{**}_{i} z^{i} \left(\frac{1}{2}\right) \overleftarrow{d}^{a} \overleftarrow{d}^{b}
\varepsilon_{ba} \right),              
\eeq
\beq
\label{E5.3}
\varepsilon( \zeta_{i} ) = \varepsilon( z^{**}_{i} ) =
\varepsilon_{i}\; ,\qquad \varepsilon( z^*_{ia} ) =  \varepsilon_{i} + 1,            
\eeq
were $\zeta_{i}$ are arbitrary external sources to $z^{i}$, while $z^*_{ia}$ , and
$z^{**}_{i}$ are the so-called
antifields to $z^{i}$,  which are, in fact, external sources to
 BRST variations of $z^{i}$ and to the composition of these BRST variations,
 respectively. Of course,
in the presence of non-zero external sources,
the  path integral (\ref{E5.1})  is in general actually dependent of gauge-fixing
Boson $F$. However, due to the
known equivalence theorem, the physical observables are gauge-independent \cite{KT}.
It is just the Ward
identity what measures the deviation
of the path integral from being  gauge-independent.

Let us perform in (\ref{E5.1})  the change $z^i\rightarrow\overline{z}^i$
of integration variables, where $\overline{z}^i$
is defined by (\ref{E4.3})  with arbitrary $\mu_a[z]$.
Then, by  using the BRST-BFV invariance (\ref{E2.16}), as well
as (\ref{E4.4}) for the Jacobian, we get what we call "a modified Ward identity",
\beq
\nonumber
&&\left< \left[ 1 + \left(\frac{i}{\hbar}\right) \int dt \zeta_{i} z^{i}
\left(\overleftarrow{d}^{a} \mu_{a} +
\left(\frac{1}{2}\right) \overleftarrow{d}^{a} \overleftarrow{d}^{b} \mu_{b} \mu_{a} \right)
+\left(\frac{i}{\hbar}\right) \int dt z^*_{ia} z^{i}
\overleftarrow{d}^{a} \overleftarrow{d}^{b} \mu_{b} +\right.\right.\\
\label{E5.4}
&&\left.\left.+(1/2)\big((i/\hbar)\!\!\int\! dt( \zeta_{i} z^{i}\overleftarrow{d}^{a} \mu_{a} +
 z^*_{ia} z^{i}\overleftarrow{d}^{a} \overleftarrow{d}^{b} \mu_{b} )\big )^{2} \right]
\left[\det(\delta_{a}^{b}+\mu_{a}\overleftarrow{d}^{b} )
 \right]^{-1}\!\! \right>_{ F; \zeta, z^*, z^{**}}\!\!\!
= 1,                  
\eeq
where we have denoted the source dependent mean value,
\beq
\label{E5.5}
&&< ( ... ) >_{F ; \zeta, z^*, z^{**}}=[ Z_{F}( \zeta, z^*, z^{**} ) ]^{-1}
\int  Dz ( ... ) \exp\left[ \left(\frac{i}{\hbar}\right)W_{F}( \zeta, z^*, z^{**})\right],\\
\nonumber
&&< 1 >_{ F; \zeta, z^*, z^{**}}  = 1,          
\eeq
related to the source dependent action in (\ref{E5.1}).
By construction, in (\ref{E5.4}) both $(\zeta_i,z^*_{ia},z^{**}_i)$
and $\mu_a$ is arbitrary. The presence  of arbitrary $\mu_a$
in the integrand in (\ref{E5.4}) reveals the implicit
dependence of the generating functional (\ref{E5.1})
on the gauge-fixing Boson $F$ for nonzero external source
$\zeta_i$.

Let us denote by $R$ the expression in the first square
bracket in the integrand in the left-hand side  in (\ref{E5.4}),
\beq \label{E5.6}
\left<R\left[\det(\delta_a^b+\mu_a[z]\overleftarrow{d}^b)\right]^{-1}\right>_{F;\zeta,z^*,z^{**}}=1.  
\eeq
By identifying in (\ref{E5.6}) the $\mu_a[z] =\mu_a[z;-\delta F]$
with the solution to the compensation
equation (\ref{E4.5}) with the inverse sign of $\delta F$,
it follows from (\ref{E5.6}) the formula generalizing (\ref{E3.4}) to
the presence of the external sources,
\beq \label{E5.7}
Z_{F_1}=Z_F<R>_{F;\zeta,z^*,z^{**}}.                           
\eeq
For a field-independent $\mu_{a} = \hbox{const}$, the latter does contribute
separately to each its order in (\ref{E5.4}), and we get from (\ref{E5.4}) to
the linear order in $\mu_a$,
\beq \label{E5.8}
\left<\int dt\left(\zeta_iz^i\overleftarrow{d}^a +
z^*_{ib}z^i\overleftarrow{d}^b\overleftarrow{d}^a\right)\right>_{F;\zeta,z^*,z^{**}}=0,       
\eeq
which is exactly the standard $Sp(2)$-form of a Ward identity.
In terms of the generating functional (\ref{E5.1}), the Ward identity
(\ref{E5.8}) is rewritten in a variation-derivative form,
\beq \label{E5.9}
\int dt\left[\zeta_i\frac{\delta}{\delta z^*_{ia}}-
\varepsilon^{ab}z^*_{ib}\frac{\delta}{\delta z^{**}_i}\right]
Z_F(\zeta,z^*,z^{**})=0.                  
\eeq

Now, let $S(z,z^*,z^{**})$ be a functional Legendre transform to
$(\hbar/i)\ln Z_F(\zeta,z^*,z^{**})$ with respect to the external source
$\zeta_i$,
\beq \label{E5.10}
z^k=\frac{\delta}{\delta \zeta_k}\left(\frac{\hbar}{i}\right)\ln Z_F(\zeta,z^*,z^{**}),                               
\eeq
\beq \label{E5.11}
S(z,z^*,z^{**})=\left(\frac{\hbar}{i}\right)\ln Z_F(\zeta,z^*,z^{**})-\int dt\zeta_iz^i,   
\eeq
\beq \label{E5.12}
S\frac{\overleftarrow{\delta}}{\delta z^i}=-\zeta_i.             
\eeq
It follows then from (\ref{E5.10})-(\ref{E5.12}) that the $Sp(2)$ master
equation,
\beq
\label{E5.13}
\left(\frac{1}{2}\right)(S,S)^a+V^aS = 0,                                     
\eeq
holds for $S$, where we have denoted the so-called "$Sp(2)$-antibracket",
\beq
\label{E5.14}
( F, G )^{a}  = \int dt\left(F\left[\frac{\overleftarrow{\delta}}{\delta z^i}
\frac{\overrightarrow{\delta}}{\delta z^*_{ia}}-
 \frac{\overleftarrow{\delta}}{\delta z^*_{ia}}
\frac{\overrightarrow{\delta}}{\delta z^i}\right]G\right)=-
(G,F)^a(-1)^{(\varepsilon_F+1)(\varepsilon_G+1)},   
\eeq
and the "$V^a$-operators",
\beq
\label{e66}
V^a=\varepsilon^{ab}\int dtz^*_{ib}\frac{\delta}{\delta z^{**}_i}   
\eeq
(for details and properties of operators used here, see \cite{BLTh1}).

\section{Discussions}

In the framework of the $Sp(2)$-extended generalized Hamiltonian
formalism \cite{BLTh1,BLTh2,BLTh3}, we have studied systematically finite
BRST-BFV transformations with two parameters being odd functionals
of phase variables. We have defined these transformation effectively
by the corresponding Lie equations in the Fermionic "plane" of the
two parameters. It was shown that the Jacobian of finite
transformation can be represented explicitly in the form of a
$2\times2$ determinant. We have formulated the $2\times2$
matrix-valued compensation equation which is sufficient to provide
for generating an arbitrary finite change of a gauge-fixing function
in the path integral. In this way, we have extended the proof of the
gauge independence of the partition function under finite variations
of gauge-fixing function. An efficient technique was developed based
on the use of the $Sp(2)$ vector-valued functional differential as
operating on the space of trajectories. It appears that all the main
objects in our consideration can be represented in a natural way in
terms
 of the functional differential
proposed. By making use of that technique as applied to finite
BRST-BFV transformation, we  derive the
$Sp(2)$ vector-valued modified Ward identities. As a particular case of the latter,
we have derived the
$Sp(2 )$ vector-valued master equation for the effective action.
As another particular case, we  have
derived the relation connecting the generating functionals
for Green functions in two arbitrary admissible gauges.

\section*{Acknowledgments}
\noindent
I. A. Batalin would like  to thank Klaus Bering of Masaryk
University for interesting discussions. The work of I. A. Batalin is
supported in part by the RFBR grants 14-01-00489 and 14-02-01171.
P. M. Lavrov thanks I. L.
Buchbinder for useful discussions. The work of P. M. Lavrov is
partially supported by the Ministry of Education and Science of
Russian Federation, grant TSPU-122, by the Presidential grant
88.2014.2 for LRSS and  by the RFBR grant 13-02-90430-Ukr.
The work of I. V. Tyutin is partially supported by the RFBR grant
14-01-00489.
\\


\begin {thebibliography}{99}
\addtolength{\itemsep}{-8pt}

\bibitem{BLT-BRST-BFV}
I.A. Batalin, P.M. Lavrov and I.V. Tyutin,
{\it A systematic study of finite BRST-BFV transformations in
generalized Hamiltonian formalism},
arXiv:1404.4154[hep-th].

\bibitem{BLT-BRST-BV}
I.A. Batalin, P.M. Lavrov and I.V. Tyutin,
{\it A systematic study of finite BRST-BV transformations in
 field-antifield formalism},
arXiv:1405.2621[hep-th].

\bibitem{FV}
E.S. Fradkin and G.A. Vilkovisky,
{\it Quantization of relativistic systems with constraints},
Phys. Lett. B55 (1975) 224.

\bibitem{BVhf}
I.A. Batalin and G.A. Vilkovisky,
{\it Relativistic $S$-matrix of dynamical systems
with boson and fermion constraints},
Phys. Lett. B69 (1977) 309.

\bibitem{FF}
E.S. Fradkin and T.E. Fradkina,
{\it Quantization of relativistic systems with Boson and Fermion
first and second class constraints},
 Phys. Lett. B72 (1978) 343.

\bibitem{BV}
I.A. Batalin and G.A. Vilkovisky,
{\it Gauge algebra and quantization},
Phys. Lett. B102 (1981) 27.

\bibitem{BV2}
I.A. Batalin  and G.A. Vilkovisky,
{\it Quantization of gauge theories
with linearly dependent generators},
Phys. Rev. D28 (1983) 2567.

\bibitem{brs1}
C. Becchi, A. Rouet and R. Stora, {\it The abelian Higgs-Kibble,
unitarity of the S-operator}, Phys. Lett. B52 (1974) 344.

\bibitem{brs2}
C. Becchi, A. Rouet and R. Stora,
 {\it Renormalization of Gauge Theories} Ann. Phys. (N. Y.)  98  (1976)
287.

\bibitem{t}
 I. V. Tyutin, {\it Gauge
invariance in field theory and statistical physics in operator
formalism}, Lebedev Institute preprint  No.  39 (1975)
(arXiv:0812.0580[hep-th]).

\bibitem{CF}
G. Curci and R. Ferrari,
{\it Slavnov transformations and supersymmetry},
Phys. Lett. B63 (1976) 91.

\bibitem{Oj}
I. Ojima,
{\it Another BRS transformation},
Prog. Theor. Phys. 64 (1979) 625.

\bibitem{Hw}
S. Hwang,
{\it Properties of the anti-BRS symmetry in a general
framework},
Nucl. Phys. B231 (1984) 386.


\bibitem{Sp}
V.P. Spiridonov,
{\it Sp(2)-covariant ghost fields in gauge theories},
Nucl. Phys. B308 (1988) 527.

\bibitem{BLTh1}
I.A. Batalin, P.M. Lavrov and I.V. Tyutin,
{\it Extended BRST quantization of gauge theories in generalized
canonical formalism},
J. Math. Phys. 31 (1990) 6.

\bibitem{BLTh2}
I.A. Batalin, P.M. Lavrov and I.V. Tyutin,
{\it An Sp(2) covariant version of generalized canonical quantization of
dynamical system with linearly dependent constraints},
 J. Math. Phys. 31 (1990) 2708.

\bibitem{BLTh3}
I.A. Batalin, P.M. Lavrov and I.V. Tyutin I.V.
{\it
An Sp(2) covariant formalism  of generalized canonical quantization of
systems with second-class constraints},
Int. J.  Mod.  Phys. 6 (1990) 3599.

\bibitem{BLT4}
I.A. Batalin, P.M. Lavrov and I.V. Tyutin,
{\it Covariant quantization of gauge theories
in the framework of extended BRST symmetry},
J.~Math. Phys. 31 (1990) 1487.

\bibitem{BLT5}
I.A. Batalin, P.M. Lavrov and I.V. Tyutin,
{\it An $Sp(2)$-covariant quantization
of gauge theories with linearly dependent generators},
J.~Math. Phys. 32 (1991) 532.

\bibitem{BLT6}
I.A. Batalin, P.M. Lavrov and I.V. Tyutin,
{\it Remarks on the $Sp(2)$-covariant quantization of gauge
theories},
J.~Math.~Phys. 32 (1991) 2513.

\bibitem{KT}
R.E.  Kallosh and I.V. Tyutin,
{\it The equivalence theorem and gauge invariance in renormalizable  theories},
Sov. J. Nucl. Phys. 17 (1973) 98.

\end{thebibliography}
\end{document}